\documentclass[12pt]{article}
\usepackage{geometry}                		% See geometry.pdf to learn the layout options. There are lots.
\usepackage{graphicx}				% Use pdf, png, jpg, or eps§ with pdflatex; use eps in DVI mode
\usepackage{amssymb}
\usepackage{bm}
\usepackage{amsmath}
\usepackage{authblk}
\usepackage{subcaption}
\usepackage{wrapfig}
\usepackage{t1enc}
\usepackage[english]{babel}
\usepackage[latin2]{inputenc}
\usepackage{graphicx}
\usepackage{babel,a4wide,amsfonts,bm}
\usepackage{slashed}
\usepackage{cite}
\usepackage{dsfont}
\usepackage{color}
\usepackage{hyperref}
\usepackage{authblk}
\def\asbar{\bar{\alpha}_s}

\newcommand{\be}{\begin{eqnarray}}
\newcommand{\ee}{\end{eqnarray}}

\newcommand{\bea}{\begin{align}}
\newcommand{\eea}{\end{align}}

\newcommand{\CK}{{\mathcal{K}}}

\newcommand{\kt}{{{\bm k}}}       % transverse momentum of exchanged gluon
\newcommand{\qt}{{\bm q}}        % transverse momentum of emitted gluon
\newcommand{\Li}{{\mathrm{Li}_2}}       % polylogarithm

\def\bea{\begin{eqnarray}}
\def\eea{\end{eqnarray}}

\def\funp{{I\!\!P}}

\graphicspath{{figs/}}

\usepackage{color}
\usepackage[normalem]{ulem}

\title{Structure functions from  renormalization group improved small $x$ evolution}

\author{Wanchen Li}
\author{Anna M. Sta\'sto}

\affil{\small \it Department of Physics, Penn State University, University Park, PA 16802, USA}

\begin{document}
\maketitle

\begin{abstract}
We perform the fit to the structure function $F_2$ data from HERA including terms due to the resummation at small $x$. The equation for the unintegrated gluon density is solved, previously established within the   renormalization group improved small $x$ framework.
We find very good description of the structure function $F_2$ and its charm component $F_2^c$. The resulting unintegrated gluon density is found to be consistent with the calculations based on similar approaches  available in the literature, with only slightly higher intercept value.
\end{abstract}

\section{Introduction}

 Deep Inelastic Scattering (DIS), the process of scattering of leptons off protons is the most precise way to explore the nucleon structure. Since the groundbreaking experiments at SLAC \cite{Breidenbach:1969kd},  which have discovered the partonic structure of the proton, enormous progress has been achieved in understanding the proton structure thanks to the series of experiments as well as theoretical developments. HERA collider in Hamburg, the only electron-proton collider up to date, operated at around $\sqrt{s}=300 \; \rm GeV$ of center-of-mass energy, and  provided important insight into the behavior of strong interactions in the small $x$ Bjorken region. One of the most important  discoveries by HERA experiments was the observation of the strong rise of the structure function \(F_2\) in the small   \(x\)  region \cite{H1:1993jmo,ZEUS:1993ppj}. This rise is understood to be driven by the sharp rise of the gluon density towards small $x$, which is a fundamental property of the strong interactions.
 
 The Dokshitzer-Gribov-Lipatov-Altarelli-Parisi (DGLAP) \cite{Altarelli:1977zs,Dokshitzer:1977sg,Gribov:1972ri,Gribov:1972rt}  evolution equation together with the collinear factorization \cite{Collins:1989gx} provide a framework for the perturbative parton density evolution. The DGLAP  evolution runs with the scale \(Q^2\)   thus it resums the powers of \( \ln\left(Q^2/Q^2_0\right)\) where $Q_0$ is some reference scale. Currently,  DGLAP splitting  functions are known up to NNLO (next-to-next-to-leading order) accuracy, and the global analyses based on the wealth of deep inelastic and hadron-collider data provide with the important information on parton densities.

The high energy framework provides an alternative approach to the parton evolution. At sufficiently high center of mass energy \(\sqrt{s}\), the \(\ln(s/s_0)\) terms will become important and one needs a different resummation approach than DGLAP. Here, $s_0$ is the energy scale choice, which for the case of DIS would correspond to $Q^2$, the minus photon virtuality. The Bjorken $x$ is then defined by $x\simeq Q^2/s$, where $s$ is the photon-proton center of mass energy squared. The resummation in the high energy limit is then accomplished by  the Balitsky-Fadin-Kuraev-Lipatov (BFKL) equation and thus accounts for the  evolution  in $\ln s/s_0$ or $\ln 1/x$ in the specific case of DIS.  The BFKL evolution equation has been derived at LL (leading logarithmic  in $\ln 1/x$) \cite{Lipatov:1985uk,Kuraev:1977fs,Balitsky:1978ic} and  NLL  \cite{Fadin:1998py,Ciafaloni:1998gs} accuracy.
The NNLL BFKL kernel has also been derived in the N=4 sYM case \cite{caron2018high,Velizhanin:2015xsa,Gromov:2015vua}. 
It has also been derived (up to NLL accuracy in QCD) in the context of the  high energy operator expansion \cite{Balitsky:1995ub,Balitsky:1998ya,Balitsky:2008zza}, dipole evolution \cite{Mueller:1989st,Mueller:1993rr,Kovchegov:1999ua,Kovchegov:1999yj} and in the Color Glass Condensate \cite{McLerran:1993ka,McLerran:1993ni,McLerran:1994vd,JalilianMarian:1996xn,JalilianMarian:1997dw,JalilianMarian:1997gr,JalilianMarian:1997jx,Iancu:2000hn,Iancu:2001ad, Kovner:2014lca,Lublinsky:2016meo}. In this framework it  includes also  high density effects, or parton saturation effects. These are  stemming from the non-linear recombination of gluons (in the context of CGC) or multiple scattering (in the context of the dipole approach) which lead to the taming of the growth towards small $x$. Together  with \(k_T\) factorization, the BFKL equation  gives prediction for  the observable quantities, like structure functions. The basic object within this framework is the unintegrated gluon density  \(f(x,k_T^2)\) which depends on the transverse momentum $k_T$ of the off-shell gluon.

The BFKL evolution at LL accuracy generates the rise of the unintegrated gluon density with decreasing $x$, with the  hard Pomeron intercept \(\omega_0 = (3\alpha_s/\pi)\, 4\ln 2\). However, it has been well known since some time that  this is too steep for the experimental data \cite{Bojak:1996nr}.

The NLL BFKL computed in \cite{Fadin:1998py,Ciafaloni:1998gs} and in \cite{Balitsky:2008zza,Kovner:2014lca,Lublinsky:2016meo} (in the context of high energy processes with saturation) turned out to be very large and negative. The result even leads to a instability, like  negative cross section, and thus hinted at the neccessity of the calculation of yet higher orders or a resummation.  The main part of the NLL corrections stem from the running of the coupling, non-singluar terms of the DGLAP splitting function and kinematical constraint. It can be shown that these contributions exhaust most  of the NLL correction \cite{Salam:1998tj,Salam:1999cn}.

In order to stabilize the results, various resummation approaches were constructed   in the literature \cite{Salam:1998tj,Salam:1999cn,Ciafaloni:1999au,Ciafaloni:1999yw,Ciafaloni:2003ek,Ciafaloni:2003rd,Ciafaloni:2003kd,Thorne:2001nr,White:2006yh,Altarelli:1999vw,Altarelli:2000mh,Altarelli:2001ji,Altarelli:2003hk,Altarelli:2008aj,Vera:2005jt,Ball:2017otu} and their results have been successfully fitted to the structure function data on $F_2$. In particular, the analysis by \cite{Ball:2017otu} demonstrated that the HERA data might actually point to the neccessity of the resummation, since the fits with the resummation resulted in slightly better $\chi^2$ than the standard fits based on the  NNLO DGLAP evolution. While not a large effect, this may have consequences for processes at yet higher energies. The resummation of low $x$ terms is especially important in the context of the next generation high energy colliders as well as for astroparticle physics. Several next generation Deep Inelastic Scattering machines are planned, like Electron Ion Collider (EIC) in the US \cite{AbdulKhalek:2021gbh}, Large Hadron electron Collider (LHeC) \cite{LHeC:2020van}, and further in the future and with yet higher energy Future Circular Collider (FCC), in the electron-hadron option \cite{FCC:2018byv,FCC:2018vvp}. The EIC machine is the project based in the US, which will utilize  the existing RHIC complex
and add the new electron accelerator. It will have variable center of mass energy up to about $\sqrt{s}=\,140\, \rm GeV$, high luminosity ${\cal L}=10^{34} \, \rm cm^{-2} s^{-1}$, and will have various nuclear targets and polarization of the colliding beams. 
The LHeC project plans to utilize the $7 \,\rm TeV$ hadron beams from the LHC and collide them with the electrons accelerated to about $50-60 \, \rm GeV$ from the Energy Recovery Linac (ERL). The center of mass energy would reach about $1\, \rm  TeV$  in this collider. The LHeC would also have high luminosity and possibility of performing electron-ion collisions with beams of lead, up to about $5 \, \rm TeV$ of center of mass energy.
The FCC in electron-hadron mode would open up another territory of the DIS. With the $50\, \rm TeV$ proton beams from the FCC, colliding with the electron beams from ERL, the center of mass energies up to $3.5 \, \rm TeV$ would be possible, as well as the collisions with the highly energetic lead beams. The precise knowledge of the evolution towards small $x$ is thus necessary for these  DIS projects as well as for the processes which are sensitive to the small $x$ dynamics in the high energy  hadronic collisions, also in the context of the astroparticle physics, see \cite{Bhattacharya:2016jce,Carpio:2020wzg}.

In this paper, we  focus on Ciafaloni-Colferai-Salam-Sta\'sto (CCSS) resummation \cite{Ciafaloni:2003ek,Ciafaloni:2003kd,Ciafaloni:2003rd}
and perform the fits to the DIS structure function data based on this framework. 
The CCSS resummation features the renormalization group improved small $x$ equation.  In the CCSS scheme the  resummation is constructed based on the  collinear splitting function and exact  BFKL up to the NLL accuracy. It contains the kinematical constraint and appropriate subtractions which avoid double counting. We find very good description of the small $x$ HERA data and the data on charm structure function. 
The CCSS resummation is formulated entirely in the momentum space, and  performing the fit to the HERA data one can then extract the resummed unintegrated gluon distribution function. The latter quantity can thus be used in other processes which make use of the $k_T$  or hybrid factorization. Similar low $x$ resummations, formulated  entirely in the momentum space \cite{Kwiecinski:1997ee,Kutak:2012rf} were successfully applied to the description of the DIS data on structure functions. In particular, we shall compare the resulting gluon distribution from CCSS to these earlier extractions \cite{Kutak:2012rf}.  We find they are consistent, but there are some differences related to the normalization  and also small differences in the evolution towards small $x$. In particular, CCSS is characterized by slightly faster growth towards small $x$. This can have implications at yet lower values of $x$, where saturation corrections (not currently included in CCSS formalism) may be needed.

The outline of this paper is as follows. In Sec. \ref{sec:CCSS}, we introduce the BFKL formalism and the CCSS resummation, recalling the formulae both in the Mellin space and in the momentum space. The details of the setup for the calculation of the structure function $F_2$ are presented in Sec.~\ref{sec:structurefunction}. In section Sec.~\ref{sec:numerical} we show the results of the fits and also extracted unintegrated gluon density.
Finally, in Sec.~\ref{sec:conclusions} we state brief conclusions and outlook.

%%%%%%%%%%%%%%%%%%%%%%%%%%%%%%%%%%%%%%%%
\section{Resummed Ciafaloni-Colferai-Salam-Stasto (CCSS) scheme}
\label{sec:CCSS}

\subsection{Recap of the BFKL evolution}
Let us first recap the basics of the BFKL evolution.
The BFKL equation has been first derived in LL approximation in the high energy limit in series of seminal papers \cite{Balitsky:1978ic,Kuraev:1977fs,Lipatov:1985uk}. It can be written generically in the following form in the momentum space
\begin{equation}
	f(x,{\bm k }) = f^{(0)}(x,{\bm{k}}) + \int_x^1 \frac{\mathrm{d}z}{z}\int \mathrm{d^2} {\bm k} \, \mathcal{K}({\bm k},{\bm k}^{\prime}) \, f(\frac{x}{z},{\bm k^{\prime }}) \; ,
	\label{BFKL integral equation}
\end{equation}
where  \(f(x,\kt)\)  is the unintegrated gluon density, which depends on fraction $x$ of the longitudinal momentum  and on the transverse momentum $\kt$ of the  reggeized gluon exchanged in the $t$ channel.
The BFKL kernel \(\mathcal{K}(\bm{k},\bm{k}_0)\) has the following perturbative expansion
\begin{equation}
	\mathcal{K}(\bm{k},\bm{k}^{\prime}) = \overline{\alpha}_s (\mu^2) \mathcal{K}_{0}(\bm{k},\bm{k}^{\prime}) \, + \, \overline{\alpha}_s^2 (\mu^2) \mathcal{K}_{1}(\bm{k},\bm{k}^{\prime}) + \, ...\; ,
	\label{momentum kernel}
\end{equation}
where the (rescaled) strong coupling constant is defined as
\be
\asbar(\mu^2) =  \frac{N_c}{\pi}\alpha_s(\mu^2) \;  ,
\ee
with $N_c$ the number of colors and   $\mu^2$ being the scale of the strong coupling. In the above equation \eqref{momentum kernel},  $\mathcal{K}_{0}$ is LL and $\mathcal{K}_{1}$ is  NLL BFKL kernel respectively. It is worth noting that, to the NLL accuracy $\mu^2$ in principle can take any form in the argument of the strong coupling the leading kernel, with appropriate subtractions in the NLL kernel to match the NLL result. On the other hand, to this level of  accuracy the scale in the coupling in  the NLL kernel is not determined. This would require the knowledge of the kernel at NNLL level.

  In the following, we shall also use the  notation $|\kt^2|=k^2$ and  work under the assumption of the   angular averaged density, i.e. $f(x,{\bm k })=f(x,k^2)$. This is justified for the inclusive case, but would be important for example in the context of the angular distributions in particle production.
It is usually convenient to study  the behavior  of the BFKL kernel in the Mellin space. The Mellin transformation is thus introduced in the following  convention, 
\begin{eqnarray}
	&&\overline{f}(\omega,k^2) = \int_0^1 \frac{\mathrm{d}x}{x}\,x^{\omega}\,f(x,k^2)\, ,\\
	&&{\mathcal{F}}(\omega,\gamma) = \int_{0}^{\infty} \mathrm{d}k^2 \left(k^2\right)^{-\gamma}\,\overline{f}(\omega,k^2)\, .
\end{eqnarray} 
In Mellin space, the BFKL equation can be rewritten into this general analytical form
\begin{equation}
{\mathcal{F}}(\omega,\gamma) = {\mathcal{F}}^{(0)}(\omega,\gamma) + \frac{\asbar}{\omega}\, \chi (\gamma)\, {\mathcal{F}}(\omega,\gamma) \, .
\end{equation}
Similar to the Eq. (\ref{momentum kernel}), the kernel \( \chi (\gamma)\) can be also computed in different orders of \(\overline{\alpha}_s\),
\begin{equation}
	\chi (\gamma) = \asbar \chi_0(\gamma) + \asbar^2 \chi_1(\gamma) + .... \, .
\end{equation}
The expressions for the LL and NLL kernels are well known in QCD and they read \cite{Lipatov:1985uk, Fadin:1998py,Ciafaloni:1998gs}
\begin{equation}
	\chi_0(\gamma) = 2\psi(1) - \psi(\gamma) - \psi (1-\gamma) \, ,
\end{equation}
\begin{eqnarray}
	\chi_1(\gamma)& =& \frac{b}{2}\left[\chi_0^2(\gamma)+\chi_0'(\gamma)\right]-\frac{1}{4}\chi_0''(\gamma) -\frac{1}{4}\left(\frac{\pi}{\sin \pi \gamma}\right)^2\frac{\cos \pi \gamma}{3(1-2\gamma)}\left[11+\frac{\gamma(1-\gamma)}{(1+2\gamma)(3-2\gamma)}\right] \nonumber\\
	&+&\left(\frac{67}{36}-\frac{\pi^2}{12}\right)\chi_0(\gamma) + \frac{3}{2}\xi(3)+\frac{\pi^2}{4\sin \pi \gamma } - \Phi(\gamma) \, .
	\label{eq:nllorg}
\end{eqnarray}
where \(\psi(\gamma)\) is polygamma function, \(b = (33-2N_f)/(12\pi)\) with \(N_f\) the number of  quarkflavors, and 
\begin{equation}
	\Phi(\gamma) = \sum_{n=0}^{\infty} (-1)^n\left[\frac{\psi(n+1+\gamma)-\psi(1)}{(n+\gamma)^2}+\frac{\psi(n+2-\gamma)-\psi(1)}{(n+1+\gamma)^2}\right].
\end{equation}
In  N=4 sYM  theory the kernel $\chi$ has been calculated up to NNLL order \cite{Velizhanin:2015xsa,Gromov:2015vua,caron2018high}.
In the leading logarithmic approximation the functions $(k'^2)^\gamma$ are the eigenfunctions of the equation, and $\chi_0(\gamma)$ is the eigenvalue at the leading logarithmic order.
This is possible since the coupling is not running and the equation is scale invariant. At NLL, due to the running of the coupling the conformal invariance of the lowest order result is broken. 
The eigenfunctions become scale dependent and include the running coupling  \cite{Chirilli:2013kca}. 
We note that, nevertheless, it is possible to analyze this equation in the Mellin space, even with the running coupling,  including higher order corrections.

 The expression in Eq.~\eqref{eq:nllorg} is given for the symmetric scale choice. In order to transform to the asymmetric scale choice, for example for $s_0=k^2$ in the Mellin space one needs to add to  Eq.~\eqref{eq:nllorg}, the term
\begin{equation}
-\frac{1}{2} \chi_0(\gamma) \frac{\partial \chi_0}{\partial \gamma}\;,
\end{equation}
which is the NLL term due to the scale changing transformation, see \cite{Ciafaloni:1998gs}.  At NNLL the scale changing transformation is much more complicated as it  involves the NLL kernel as well, as was shown in \cite{Marzani:2007gk,Deak:2019wms}.

 It has been known since some time that NLL corrections are numerically very large \cite{Fadin:1998py} and lead to the  instabilities in the solution, including possibility of oscillating and negative cross sections \cite{Salam:1998tj}.  This  has also been seen in the case of the NLL corrections to the non-linear equation, \cite{Lappi:2015fma}, meaning in particular that the saturation corrections do not reduce the effect of the NLL terms.
The problem has been identified early on, as originating from the  double and triple collinear poles in the Mellin space $\gamma$. These are poles are coming from the additional transverse logarithms in the momentum space \cite{Salam:1998tj,Salam:1999cn}.  The collinear poles, which correspond to the strong ordering when $k^2 \gg k'^2$ appear as $1/\gamma$ and the anti-collinear poles which correspond to $k^2 \ll k'^2$ as $1/(1-\gamma)$.  
At NLL level in the high energy limit the double  poles arise due to the non-singular (in $1/z$) part of the DGLAP splitting function as well as due to the running of the coupling. In addition there are also 
 triple collinear poles which appear due to the kinematical constraints \cite{Ciafaloni:2003ek,Ciafaloni:2003rd}. 
 The fact that the BFKL equation receives large corrections from the kinematics was identified early on, before the full NLL calculation.
The kinematical constraint \cite{Ciafaloni:1987ur} was shown to  originate from the improved kinematics, and more precisely from the requirement that the exchanged momenta are dominated by the transverse components \cite{Andersson:1995ju,Kwiecinski:1996td}. There are different forms of the kinematical constraint  available in the literature, however they all resum the triple collinear poles  at NLL and higher at more subleading orders.   For more recent work on different forms of kinematical constraint see \cite{Deak:2019wms}. There it  has been shown that kinematical constraint is consistent with the NNLL calculation in N=4 sYM, and the resummation of the $1/\gamma^5$ poles  was demonstrated.

%%%%%%%%%%%%%%%%%%%%%%%%%%%%%%%%%%%%%%%%%%%%
\subsection{The CCSS resummed equation}
\label{sec:ccss_kernel}

The CCSS resummation scheme was formulated in \cite{Ciafaloni:2003ek}, where more details can be found. Here we only summarize main points and features of the construction, essential for the subsequent discussion. The resummation was motivated in the Mellin space, where the analysis of the collinear poles is more transparent. The final formulation however was constructed in the momentum space, and the integral equation for the gluon Green's function solved, see \cite{Ciafaloni:2003ek} and more recently \cite{Deak:2020zay}. One of the advantages of the formulating the resummation in the momentum space in the form of the integral equation, is the fact that the incorporation of the running of the coupling is more straightforward. Also, one can then solve directly for the unintegrated gluon density, which  can then be used for other phenomenological applications. This is the approach which we shall adopt in this work too.   A similar  idea for the resummation was formulated previously  in Ref.~\cite{Kwiecinski:1997ee} by combining the DGLAP and BFKL evolution with the kinematical constraint. The latter approach was formulated  without full NLL BFKL correction.  A  fit in the same spirit was performed in \cite{Kutak:2012rf}, together with the nonlinear term, and we shall perform more detailed comparison of  the results of the CCSS resummation fits with that approach in   Sec.~\ref{sec:numerical}.

In the CCSS scheme the starting point is the fixed order LL + NLL calculation. On top of that, the kinematical constraint as well as the non-singular part of the DGLAP  splitting function are added. Next, appropriate subtractions are made from the NLL kernel, which remove the double and triple poles and which are already incorporated in the kinematical constraints and DGLAP terms. This is to avoid the double counting. Finally, the $\beta_0$ function dependent term is subtracted from the NLL part and included in the running of the coupling  in front of the leading logarithmic kernel. The scale was taken equal to be 
the transverse momentum of the emitted gluon. In that way there are no terms proportional to the beta function in the NLL kernel.
The expression for the  NLL part of the resummed kernel in the Mellin space is then
\begin{equation}
\chi_{1}^{\rm subtr}(\gamma) \; = \;  \chi_1(\gamma)
+\frac{1}{2} \chi_0(\gamma)  \frac{\pi^2}{\sin^2(\pi\gamma)}-\chi_0(\gamma) \frac{A_1(0)}{\gamma(1-\gamma)}+\frac{b}{2} (\chi_0'+\chi_0^2) \;.
\label{eq:subtractions}
\end{equation}
The first term is the fixed order NLL term, the same as in Eq.\eqref{eq:nllorg}. The next three terms are the subtractions: due to  the kinematical constraint,  DGLAP terms and the running coupling.
In addition, more subtractions were added  to avoid the double counting and ensure the conservation of the momentum sum rule. Certain flexibility as to how to incorporate such subtractions leads to the various resummation schemes.

 As already mentioned above, the choice of the scale for the coupling constant in the NLL kernel is arbitrary. This is because any change at this level would require the knowledge of the NNLL term, which is currently not known in QCD.  In the CCSS scheme \cite{Ciafaloni:2003ek} the  scale at NLL was chosen to be the maximum of  the momenta squared of the exchanged gluons, i.e. $\max(k^2,k^{\prime 2})$. We will also adopt this choice for our calculation.

 The final result of the CCSS resummation was formulated directly in the momentum space through integral equation. As mentioned above, this allows for more control over the implementation of the running coupling corrections. Together with the DGLAP splitting functions this equation becomes an integral equation in both the longitudinal $z$ and transverse momentum components ${\kt}$. The three main contributions
 
 \begin{equation}
\CK_0^{\rm kc}(z;\kt,\kt') \stackrel{z,\qt}{\otimes} f(\frac{x}{z},k')+\CK_0^{\rm coll}(z;k,k') \stackrel{z,k'}{\otimes} f(\frac{x}{z},k')+\CK_1^{\rm subtr}(z;k,k') \stackrel{z,k'}{\otimes} f(\frac{x}{z},k') \; ,
\label{eq:kterms}
\end{equation}
 are coming from the leading logarithmic kernel with kinematical constraint, the so-called collinear (DGLAP) part and the NLL part with subtractions.
 
 The first term in Eq.~\eqref{eq:kterms} is 
\begin{multline}
\CK_0^{\rm kc}(z;\kt,\kt') \stackrel{z,\qt}{\otimes} f(\frac{x}{z},k') \\
 = \int_x^1 \frac{dz}{z} \int \frac{d^2 \qt}{\pi \qt^2} \;
  \bar{\alpha}_s(\qt^2) \left[ f(\frac{x}{z},|\kt+\qt|)
 \Theta(\frac{k^2}{z}-k^{\prime 2})-\Theta(k-q) f(\frac{x}{z},k) \right]\;,
 \label{eq:LOBFKLkc}
\end{multline}
where $\qt = \kt-\kt'$ corresponds to the transverse momentum  of the emitted gluon.  This choice  of a scale in the running coupling is convenient since in this case the $b$ dependent terms in the NLL part of the kernel are exactly zero.
The kinematical (or consistency) constraint is implemented onto the real emissions only (see discussion in \cite{Kwiecinski:1996td} and \cite{Deak:2020zay}). It is here asymmetric, which corresponds to the asymmetric scale choice suitable for the DIS problem we are considering.
It is implemented as
\begin{equation}
k^{\prime 2} \le \frac{k^2}{z} \; .
\end{equation}
To be precise, there are different versions of this constraint which appear in the literature, see \cite{Andersson:1995ju,Kwiecinski:1996td,Ciafaloni:1987ur}. Detailed analysis (see \cite{Deak:2019wms}) showed that all versions are generating the same leading $1/\gamma^3$ poles in the Mellin space at NLL, $1/\gamma^5$ poles in NNLL level (for supersymmetric case) and they do not generate any double poles, and with the difference starting to appear in the single pole level.

The second contribution in \eqref{eq:kterms}  is 
\begin{multline}
 \CK_0^{\rm coll}(z;k,k') \stackrel{z,k'}{\otimes} f(\frac{x}{z},k')= 
  \int_x^1 { dz \over z} \int_{0}^{k^2} \frac{{dk'}^2}{k^2} \;
  \bar{\alpha}_s(k^2) z\tilde{P}_{gg}(z)
  f(\frac{x}{z},k')   \\
 + \int_x^1 { dz \over z} \int_{k^2}^{k^2/z} \frac{{dk'}^2}{k'{}^2} \;
  \bar{\alpha}_s({k'}^2) z{{k^{\prime 2}} \over k^2} \tilde{P}_{gg}(z{k^{\prime 2} \over k^2})
  f(\frac{x}{z},k') \;.  
  \label{eq:dglapterms}
\end{multline}
It is the sum of the collinear and anticollinear parts with the non-singular part of the splitting function
\begin{equation}
 \tilde{P}^{(0)}_{gg}  =  P^{(0)}_{gg} - { 1 \over z} \;,
\end{equation}
where the $ P^{(0)}_{gg}$ is the   DGLAP gluon-gluon splitting function in LO.

Finally, the last  term  in Eq.\eqref{eq:kterms} is the  NLL part of the BFKL with appropriate subtractions  (corresponding to expression in Eq.~\eqref{eq:subtractions})  transformed into momentum space
\begin{align}
 \int_x^1\frac{dz}{z} & \int d k'{}^2 \;   \bar{\alpha}^2_s({k}^2_{>})
  \tilde{K}_1(k,k') f(\frac{x}{z},k')  \; = \;  
   {1\over 4}\int_x^1 \frac{dz}{z} \int d{k'}^2 \; \bar{\alpha}^2_s({k}^2_{>})
  \bigg\{  \nonumber \\
& {\left({67 \over 9} - {\pi^2 \over 3}\right) {1\over |{k'}^2-k^2|}
  \left [f(\frac{x}{z},{k'}^2) - {2 k_{<}^2 \over ({k'}^2 + k^2)}
  f(\frac{x}{z},k^2)\right] + } \nonumber \\
& {\bigg[ - {1 \over 32} \left({2 \over {k'}^2} + {2 \over k^2} +
  \left({ 1\over {k'}^2 } - {1\over k^2} \right)
  \log\left({k^2 \over {k'}^2}\right)\right)
  + {4 \Li(1-k_{<}^2/k_{>}^2) \over |{k'}^2 - k^2|}} \nonumber \\
 &{-4 A_1(0){\rm sgn}({k}^2-{k'}^2)
  \left( {1 \over k^2} \log{|{k'}^2-k^2| \over {k'}^2} -
  {1 \over {k'}^2} \log{|{k'}^2-k^2| \over {k}^2}\right)} \nonumber \\
& - \left(3 + \left({3 \over 4} - {({k'}^2+k^2)^2 \over 32{k'}^2 k^2}\right)
  \right) \int_0^{\infty} {dy \over k^2 + y^2 {k'}^2 }
  \log|{1+y \over 1-y}| \nonumber \\
 &+ {1 \over {k'}^2 + k^2} \left( {\pi^2 \over 3} +
  4 \Li( {k_{<}^2 \over k_{>}^2})\right) \bigg]
  f(\frac{x}{z},k') \bigg\} \nonumber \\
& + {1\over 4} 6 \zeta(3) \int_x^1 \frac{dz}{z} \;
  \bar{\alpha}^2_s(k^2)  f(\frac{x}{z},k) \;. \hspace{40mm}
\end{align}

 The above construction for the resummed kernel needs to be supplemented by additional subtractions. It turns out, \cite{Ciafaloni:2003ek} that there are terms which are giving 
 spurious DGLAP anomalous dimension at NLO. This needs to be canceled by appropriate subtraction and  
 it was achieved  by adding extra terms to the kernel.  Obviously, there is some ambiguity in this  procedure, since one is working with the information up to a fixed order in perturbation theory. Therefore  two different schemes were proposed $A$,$B$ in \cite{Ciafaloni:2003ek}. In the following, we shall utilize scheme $B$ from that work.

%%%%%%%%%%%%%%%%%%
\section{Contributions to structure function}
\label{sec:structurefunction}

The structure function \(F_2\) can be evaluated by the \(k_T\) factorization theorem, which involves an off-shell matrix element and the unintegrated gluon density. 
The structure function \( F_2\) receives however large contributions from the non-perturbative, or soft, regime. This is parametrized in our description as the contribution coming from the low momenta of the gluon $k^2$ and with the addition of the soft Pomeron contribution. The setup is similar  to the one presented in \cite{Kwiecinski:1996td}, without however the matrix formulation which would involve the evolution of quarks.

%%%%%%%%%%%%%%%%%%%%%%
\subsection{Perturbative contribution}

The perturbative contribution to the structure function is based on the \(k_T\) factorization theorem, together with the unintegrated gluon density obtained from the CCSS resummed evolution equation discussed in the previous section.
The expression for the structure function \(F_2\) from the $k_T$ factorization  is given by
\begin{equation}
	F_2(x,Q^2) = \sum_q \, e_q^2 \,  S_q(x,Q^2) \; ,
\end{equation}
where the sum is over the quark flavors and general expression for \( S_q(x,Q^2)\) is
\begin{equation}
	S_q (x, Q^2) \; = \; \int_x^1 \: \frac{dz}{z} \: \int
	\: \frac{dk^2}{k^2} \: S_{\rm box}^q \: (z, m_q^2,k^2, Q^2) \: f \left
	(
	\frac{x}{z}, k^2 \right ) \; .
	\label{eq:sqbox}
\end{equation}
The explicit expression for the convolution of the matrix element with the unintegrated gluon density is given by \cite{Askew:1992tw, Kwiecinski:1997ee}
\begin{align}
	\label{eq:boxc}
	S_q(x,Q^2) & = &  \frac{Q^2}{4\pi^2}\int \frac{dk^2}{k^4}\int_{0}^{1}d\beta\int d \kappa' \alpha_s \left\{\left[\beta^2+(1-\beta^2)\right]\left(\frac{\bm{\kappa}}{D_{1q}}-\frac{\bm{\kappa}-\bm{k}}{D_{2q}}\right)^2\right. \nonumber \\
	& + & \left.\left[m_q^2 + 4Q^2\beta^2 (1-\beta)^2\right]\left(\frac{1}{D_{1q}}-\frac{1}{D_{2q}}\right)^2\right\} f\left(\frac{x}{z},k^2\right)\Theta\left(1-\frac{x}{z}\right).
\end{align}
In the above, \( \bm{\kappa}\) and \(\bm{k}\) are quark and gluon transverse momenta respectively, and $\beta$ is the variable defined in the Sudakov decomposition of the quark momentum (longitudinal momentum fraction of the photon carried by the quark, for details see Ref.\cite{Askew:1992tw}).
In addition it is useful to defined the shifted quark transverse momentum is \(\bm{\kappa}'=\bm{\kappa} -(1-\beta)\bm{k}\).   The energy denominators are
\begin{eqnarray}
    	D_{1q} & = &\kappa^2+\beta(1-\beta)Q^2+m_q^2 \; ,\\
	D_{2q} & = &(\bm{\kappa}-\bm{k})^2+\beta(1-\beta)Q^2+m_q^2 \; .
\end{eqnarray}

The argument of the unintegrated gluon density is equal to $x/z$ with
\begin{equation}
    	z  = \left[1+\frac{\kappa'^2+m_q^2}{\beta (1-\beta) Q^2}+\frac{k^2}{Q^2}\right]^{-1} \; .
\end{equation}
This stems from the exact kinematics in the photon-gluon fusion process, see \cite{Askew:1992tw}. As analyzed in detail in \cite{Bialas:2000xs,Bialas:2001ks} the exact kinematics in the impact factor, goes beyond the leading order approximation in high energy. It has been demonstrated that it leads to large  effect numerically and is important for phenomenology \cite{Golec-Biernat:2009mod}. 

The argument of the strong coupling \(\alpha_s\)  is taken to be \((k^2+\kappa^2+m_q^2)\) in this analysis.  The masses of quarks are taken to be
 \(m_u=m_d=m_s=0\) and \(m_c=1.4\ \text{GeV}\). 
The integration over the transverse momenta in the $k_T$ factorization formula formally extends down to zero into the non-perturbative region. We assume the validity  of the formula \eqref{eq:sqbox} and \eqref{eq:boxc} only for the transverse momenta \(k^2, \kappa^2 > k_0^2\) where cutoff $k_0^2$ parametrizes  the boundary between the   perturbative and non-perturbative regions of the transverse momentum.
We took the value of \( k_0^2 = 1. \, \rm GeV^2\) for the cutoff.

%%%%%%%%%%%%%%%%%%%%%%%%%%%%%%%%%%%%%%%%%%%%%%%%%%%%%%%
\subsection{Non-perturbative contribution}

The  structure function $F_2$ receives large soft contribution. For example in the approach of \cite{Askew:1993jk}, it has been simply parametrized as the constant background term in addition to the perturbative small $x$ part. In the approaches within the dipole model, the non-perturbative contribution is usually taken automatically into account by integration over the large dipole sizes with the flat dipole cross section, see for example discussion in \cite{Berger:2011ew,Mantysaari:2018zdd}. Here we follow the approach of \cite{Kwiecinski:1997ee} where the non-perturbative contribution from the  low gluon and quark transverse momenta can be parametrized as follows. 
In the non-perturbative region, where both quark momenta and gluon momenta are small \(k^2, \kappa'^2 < k_0^2\), we assume that light  quark contribution  is phenomenologically evaluated as the soft Pomeron exchange \cite{Donnachie:1992ny}. That is we assume the soft Pomeron form for $u,d,s$ contributions
\begin{equation}
	S^{(a)} \; = \; S_u^\funp \: + \: S_d^\funp \: + \: S_s^\funp \; ,
	\label{eq:soft0}
\end{equation}
and
\begin{equation}
		S_u^\funp \; = \; S_d^\funp \; = \; 2S_s^\funp \; = \; C_\funp \:
	x^{-\lambda} \: (1 - x)^8,
	\label{eq:soft}
\end{equation}
where coefficient \( C_\funp\) is a free parameter independent of \(Q^2\). Here, $0 \lesssim \lambda \lesssim 0.1$ is the soft-Pomeron power.

In the region where the quark momenta are higher but the gluon momenta are very low, i.e.
 \(k^2 < k_0^2 < \kappa'^{ 2}\), we assume the  strong-ordered approximation at quark-gluon vertex and  used the the collinear approximation 
\begin{equation}
		S_{\rm box} \; \rightarrow \; S_{\rm box}^{(b)} \:
	(z, k^2 = 0, Q^2),
\end{equation}
and thus we have
\begin{eqnarray}
	S^{(b)} & = & \int_x^1 \: \frac{\mathrm{d}z}{z} \: S_{\rm box}^{(b)} \:
(z,
k^2 = 0, Q^2) \: \int_0^{k_0^2} \: \frac{\mathrm{d}k^2}{k^2} \:
f \left ( \frac{x}{z}, k^2 \right
) \nonumber \\
& = & \int_x^1 \: \frac{\mathrm{d}z}{z} \: S_{\rm box}^{(b)}
\: (z, k^2 = 0, Q^2) \: \frac{x}{z} \: g
\left ( \frac{x}{z}, k_0^2 \right )  \; ,
\label{eq:sb}
\end{eqnarray}
where \( x  g
\left ( x, k_0^2 \right )  \) is the non-perturbative input collinear gluon density  at scale $k_0^2$, and the form of its parametrization will be specified in Sec.~\ref{sec:numerical}.

Thus the complete contribution from the light quarks is simply the sum of 
\begin{equation}
S^{(a)}_{q}+S^{(b)}_{q}+S^{(c)}_{q} \; ,
\label{eq:sqbox_sum}
\end{equation}
where  the last, perturbative contribution is computed from Eq.~\eqref{eq:boxc} with the lower bound on the transverse momentum integration given by the cutoff $k_0^2$.

%%%%%%%%%%%%%%%%%%%%%%%%%%%%%%%%%%%%%%%%%%%%%%%%%%%%%%%%
\subsection{Charm quark contribution}

In addition to the light quarks, one needs to include the charm quark contribution, which is also evaluated from the $k_T$ factorization. The mass of the charm quark has been taken to be $m_c=1.4\;\rm GeV$.

For the gluon transverse momenta larger than the cutoff \(k^2>k_0^2\), we treat the charm contributions perturbatively, using Eq.\eqref{eq:boxc}. In the region \(k^2 <k_0^2\), we use on-shell approximation, i.e. \(k^2=0\). Therefore the contribution is then
\begin{eqnarray}
    	S^{(b)}_c & =&  \int_x^a \frac{\mathrm{d}z}{z} S_{\text{box}}(z,k^2=0,Q^2;m_c^2)\int_0^{k_0^2}\frac{\mathrm{d}k^2}{k^2}f\left ( \frac{x}{z}, k_0^2 \right )  \nonumber \\
& =	&\int_x^a \frac{\mathrm{d}z}{z} S_{\text{box}}(z,k^2=0,Q^2;m_c^2) \frac{x}{z} g\left ( \frac{x}{z}, k_0^2 \right ) \; ,
\end{eqnarray}
where \(a=(1+4m_c^2/Q^2)^{-1}\) and from Ref. \cite{Gluck:1994uf}
\begin{eqnarray}
	S_{\text{box}}(z,k^2=0,Q^2;m_c^2) & = & \frac{z\alpha_s}{2\pi}\left\{\ln \frac{1+\xi}{1-\xi}\left[z^2+(1-z)^2+z(1-3z)\frac{4m_c^2}{Q^2}-z^2\frac{8m_c^4}{Q^4}\right]\right. \nonumber \\
	&& \left. +  \xi \left[-1+8z(1-z)-z(1-z)\frac{4m_c^2}{Q^2}\right]\right\}.	
\end{eqnarray}
Here, \(\xi^2 = 1-(4m_c^2/Q^2)z(1-z)^{-1}\). These are two contributions from the charm quark that are included, which are dynamically generated from the the photon-gluon fusion. We do not consider any additional `intrinsic' charm contribution.

%%%%%%%%%%%%%%%%%%%%%%%%%%%%%%%%%%%%%%%%%%%%%%
\section{Numerical results}
\label{sec:numerical}
In this section we present the results of the fits to the structure functions and the properties of the extracted unintegrated gluon distribution function. The fits have been performed to the DIS HERA \(F_2\)\cite{Aaron:2009aa} and charm data \(F_2^c\) \cite{Aaron:2011gp}.

In this analysis we focus on high energy, or low-\(x\) physics,  therefore we impose the cuts on the data: \(Q^2 > 2\ \text{GeV}^2, \ x <0.01\).  Since we are working with the data on the reduced cross section, in order to exclude the  contribution from the longitudinal structure function \(F_L\), we also impose the cut on inelasticity \(y<0.6\). 

The  fitted parameters stem from the input gluon distribution and the soft  Pomeron contribution.
The initial condition for the evolution of the unintegrated gluon density from the CCSS evolution is given by the convolution of the integrated gluon density with the DGLAP splitting function as in \cite{Kwiecinski:1997ee}
\begin{equation}
\tilde{f}^{(0)} (x, k^2) \; = \; 
\frac{\alpha_S
(k^2)}{2\pi} \int_x^1 dz P_{gg} (z) \frac{x}{z} g
\left(\frac{x}{z}, k_0^2 \right) \; ,
    \label{eq:finitial}
\end{equation}
where $x g
\left(x, k_0^2 \right)$ is the integrated gluon density at scale $k_0^2$. This is done to avoid explicit parametrization of the initial term for the unintegrated gluon density in the non-perturbative region $k^2<k_0^2$. That is the input in the resummed renormalization group improved equation is of the same form as in the conventional DGLAP evolution. In the case of the unified DGLAP and BFKL evolution it proved to be a very successful approach and the input could be parametrized in a very economical form with only few free parameters \cite{Kwiecinski:1997ee}.
 For the integrated input gluon distribution we consider the form 
\begin{equation}
	xg(x,k_0^2 )=N(1-x)^{\beta} \left[ 1 + D \left( x + \epsilon \right)^\alpha \right] \; , \label{eq:inputgluon}
\end{equation}
where \(\epsilon\) is manually set to be a small positive number, as such we can prevent a potential negativity of the gluon input function when \(D < 0\) and \(\alpha<0\).

We have assumed two different  fitting scenarios. In the first one, the fit has  been performed to  both $F_2$ and $F_2^c$ data simultaneously. In the second scenario, only   $F_2$  data have been fitted, thus leaving the charm structure function $F_2^c$ as a prediction. The resulting parameters of the fit, which include the parameters on the initial gluon distribution and the parameters of the soft Pomeron  part, \eqref{eq:soft}, are shown in Table \ref{table:parameters}. We observe that the quality of the fits is very good and  similar in both cases, as are also the resulting values of the parameters, which demonstrates the stability of the approach.

\begin{figure}[!htb]
	\centering
	\includegraphics[width=\linewidth]{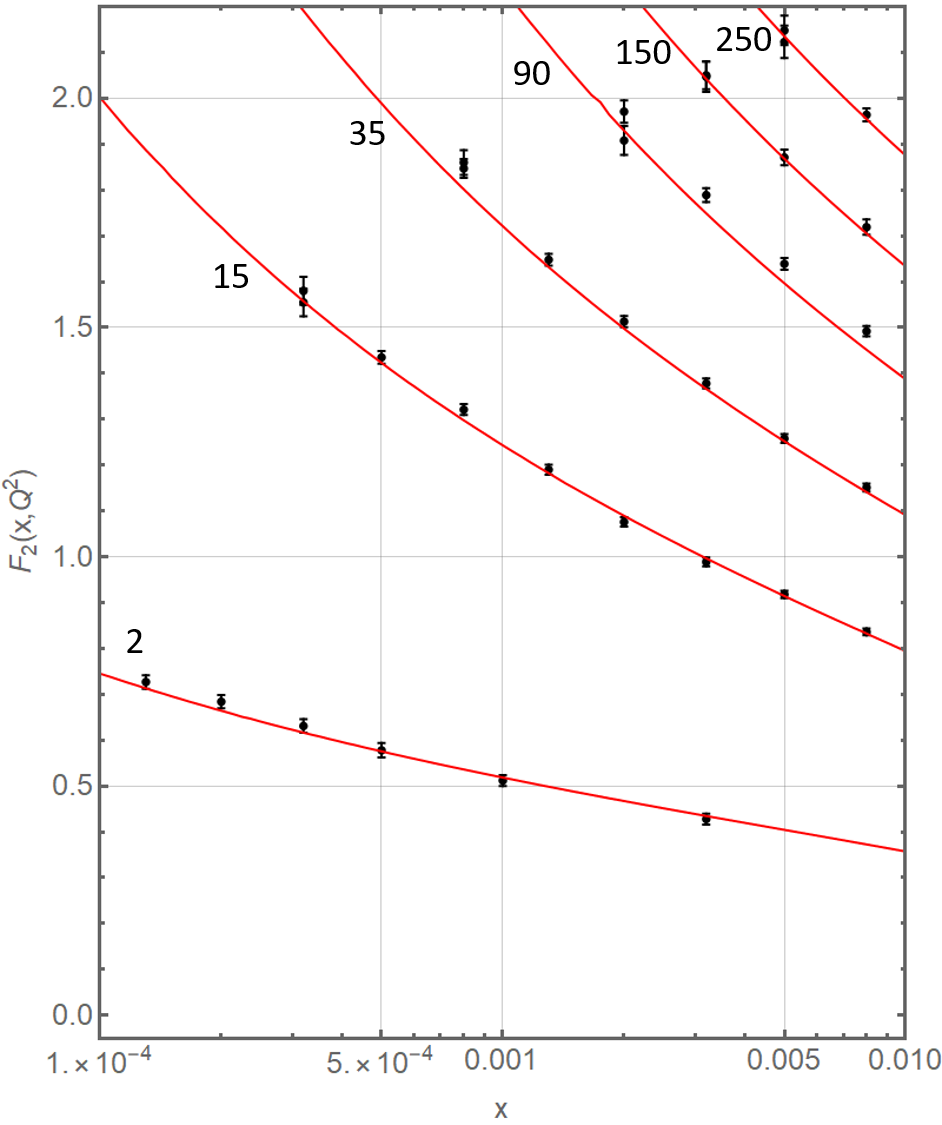}
	\caption{Structure function $F_2(x,Q^2)$  as a function of  $x$ for fixed values of $Q^2=2,15,35,90,150,250\,\rm GeV^2$, indicated next to the curves. Solid red lines  correspond to a fit with the CCSS resummed scheme. Experimental data are from Ref.~\cite{Aaron:2009aa}. }
	\label{fig:F2plot}
\end{figure}

\begin{figure}[!htb]
	\centering
	\includegraphics[width=\linewidth]{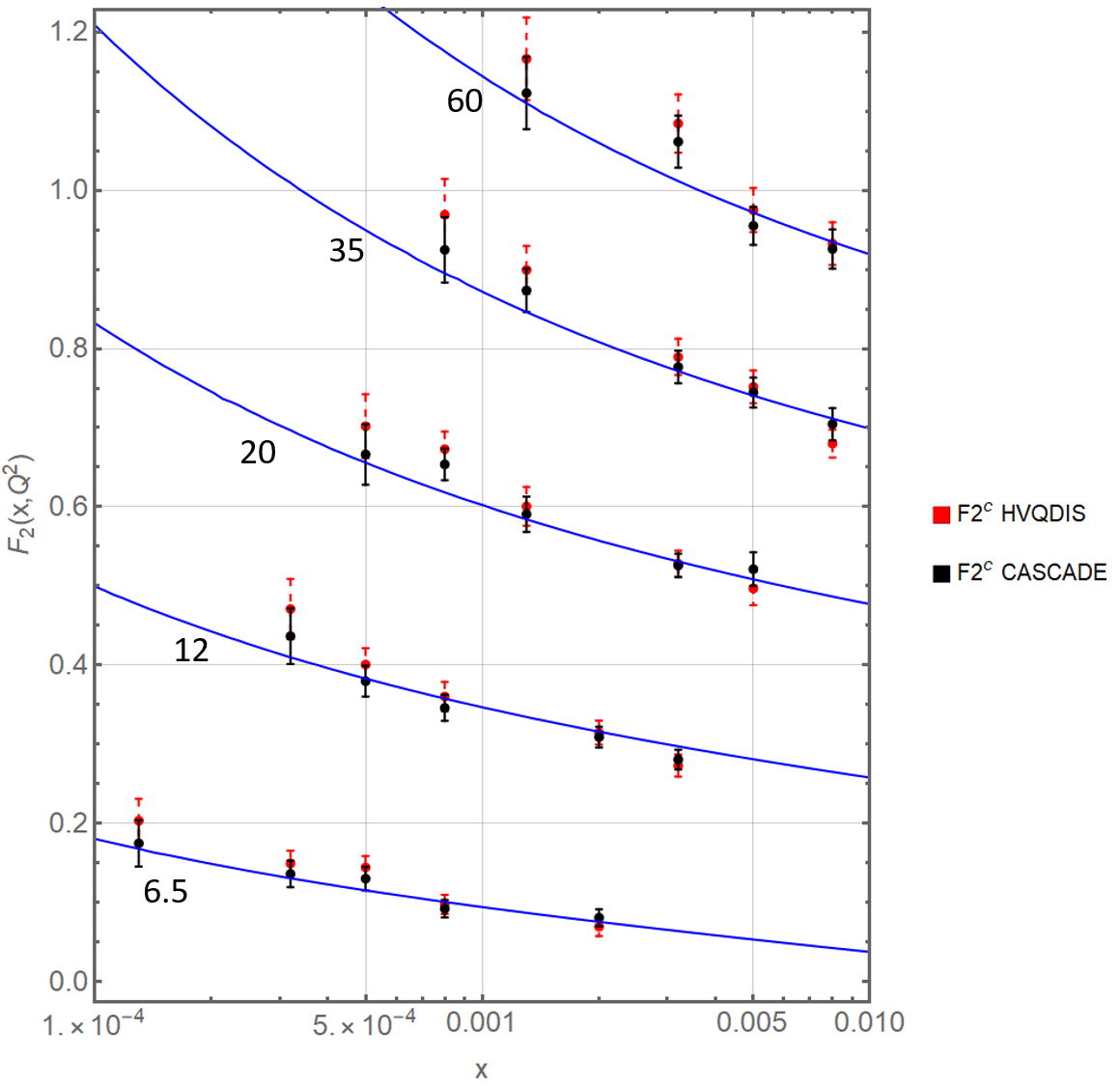}
	\caption{Charm structure function $F_2^c(x,Q^2)$  as a function of  $x$ for fixed values of $Q^2=6.5,12,20,35,60\,\rm GeV^2$, indicated next to the curves. Solid blue  curves indicates a fit using the CCSS resummed scheme.  The experimental data  using different phase space extrapolations based on theoretical calculations  CASCADE and HVQDIS are  from   Ref.~\cite{Aaron:2011gp}.}
	\label{fig:F2charmplot}
\end{figure}

\begin{table}[!htb]
	\caption{Fitting parameters for the gluon distribution and the soft Pomeron part.}
		\centering
		\begin{tabular}{c|c|c|c|c|c|c|c|c}
			Data Range & \(\chi^2\) & \(C_\funp\)& \(\lambda\) &	\(N\)&	\(\beta\)&	\(\alpha\) &	\(D\) 	&	\(\epsilon\)\\
			\hline
			Fit $F_2$,$F_2^c$ & 0.9900 & 0.4420 & 0.01562 & 3.894 & 4.951 & -0.4402 & -0.1096 & 0.0003\\
			Fit $F_2$ & 1.052 & 0.4427 & 0.01557 & 3.887 & 4.950 & -0.4413 & -0.1084 & 0.0003
		\end{tabular}
	\label{parameters}
	\label{table:parameters}
\end{table}
In the following we thus only show the results obtained when the charm structure function is also fitted.
In Fig.~\ref{fig:F2plot} we show the structure function $F_2(x,Q^2)$ plotted as a function of $x$ for  selected values of $Q^2=2,15,35,90,150,250 \; \rm GeV^2$. We see that the calculation describes the experimental data on structure function $F_2$ very well. In Fig.~\ref{fig:F2charmplot} the calculation is shown for the charm structure function $F_2^c$. The red points indicate the  charm component extracted from experiment assuming extrapolating factors to the full phase space in the HVQDIS scenario and black points in the CASCADE scheme, for details see \cite{Aaron:2011gp}.
We see again that the description of the experimental data on charm structure function is very good.
As mentioned earlier, the quality of the description is comparable to the case when charm is not fitted but is a prediction.

\begin{figure}[!htb]
	\centering
	\includegraphics[width=\linewidth]{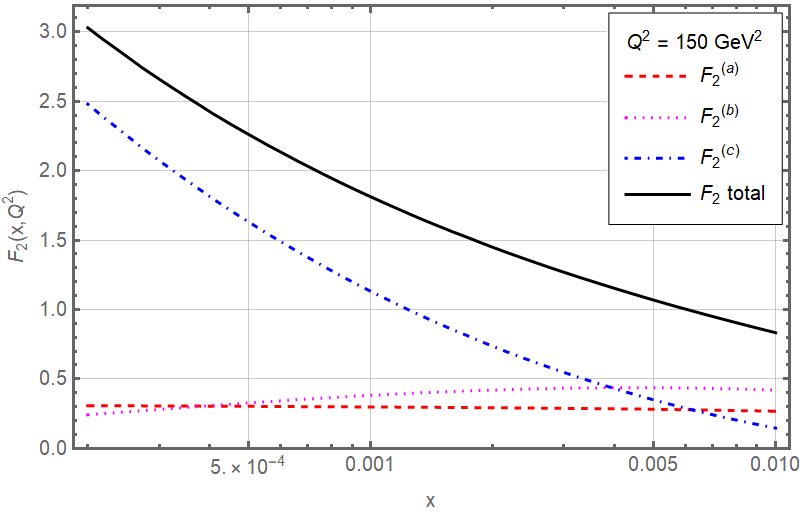}
	\caption{Structure function $F_2(x,Q^2)$ as a function of $x$ for fixed value of $Q^2=150\, \rm GeV^2$ broken down into various contributions. Red dashed: $F_2^{(a)}$,  Eq.~\eqref{eq:soft0}; pink dotted $F_2^{(b)}$, Eq.~\eqref{eq:sb}; blue dashed-dotted $F_2^{(c)}$, Eq.~\eqref{eq:sqbox}. Finally,  black solid indicates  the sum of all contributions.}
	\label{fig:F2abcQ150}
\end{figure}

\begin{figure}[!htb]
	\centering
	\includegraphics[width=\linewidth]{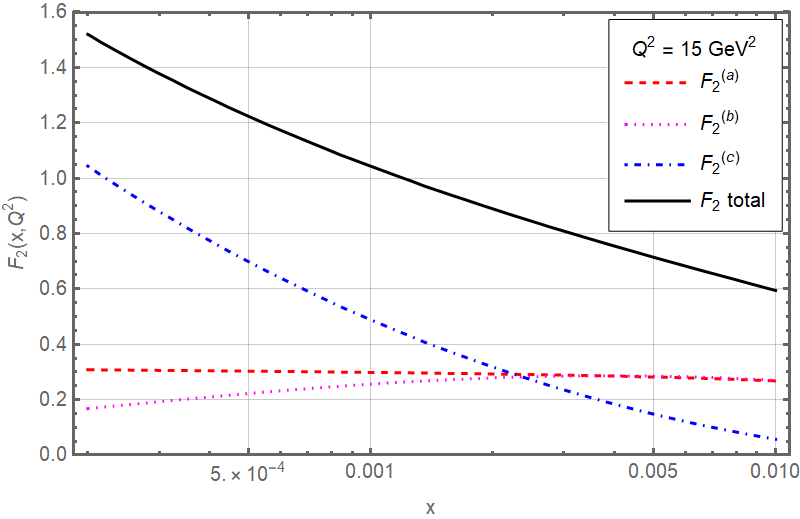}
	\caption{Structure function $F_2(x,Q^2)$ as a function of $x$ for fixed value of $Q^2=15\, \rm GeV^2$ broken down into various contributions. Red dashed: $F_2^{(a)}$,  Eq.~\eqref{eq:soft0}; pink dotted $F_2^{(b)}$, Eq.~\eqref{eq:sb}; blue dashed-dotted $F_2^{(c)}$, Eq.~\eqref{eq:sqbox}. Finally,  black solid indicates  the sum of all contributions. }
	\label{fig:F2abcQ15}
\end{figure}

\begin{figure}[!htb]
	\centering
	\includegraphics[width=\linewidth]{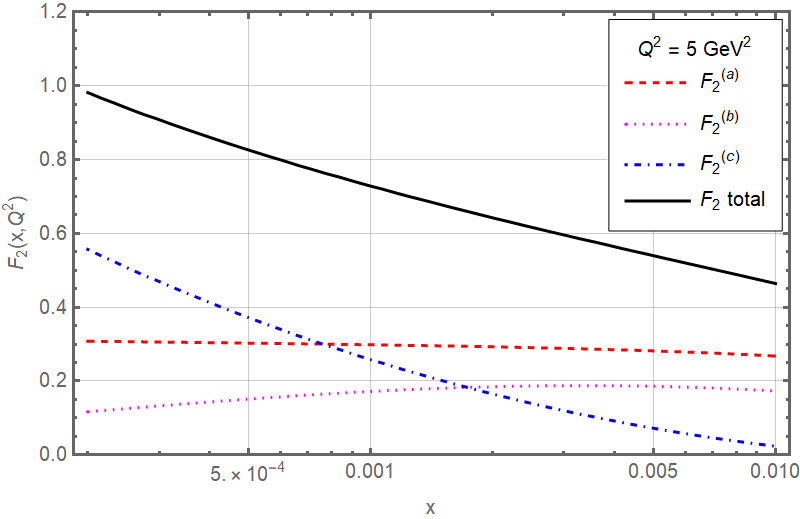}
	\caption{Structure function $F_2(x,Q^2)$ as a function of $x$ for fixed value of $Q^2=5\, \rm GeV^2$ broken down into various contributions. Red dashed: $F_2^{(a)}$,  Eq.~\eqref{eq:soft0}; pink dotted $F_2^{(b)}$, Eq.~\eqref{eq:sb}; blue dashed-dotted $F_2^{(c)}$, Eq.~\eqref{eq:sqbox}. Finally,  black solid indicates  the sum of all contributions. }
	\label{fig:F2abcQ5}
\end{figure}

In Figs.~\ref{fig:F2abcQ150},\ref{fig:F2abcQ15},\ref{fig:F2abcQ5}
we show the structure function broken down into separate components $F_2^{(a)}, F_2^{(b)}, F_2^{(c)}$ as a function of $x$ for three values of $Q^2=150, 15, 5\; \rm GeV^2$ respectively.

The perturbative component, indicated as $F_2^{(c)}$ in  figures  \ref{fig:F2abcQ15} and \ref{fig:F2abcQ150} dominates at large values of $Q^2$ and small $x$. The two  non-perturbative components have very flat dependence on $x$ and contribute to the most of the cross section  at moderate $x$ and lower $Q^2$. We see that the non-perturbative contribution due to the soft Pomeron is substantial. For example, it is  about $25\%$  at small $x \simeq 5\times 10^{-4}$ and   $Q^2 = 15 \; \rm GeV^2$. Even at high $Q^2$ the soft component is still non-negligible.  In the low \(Q^2\) region,  see Fig.~\ref{fig:F2abcQ5},  the non-perturbative components dominate the structure function in a very wide range of $x$, down to $x\sim 3\times 10^{-4}$.
It is only at smallest $x$ where the perturbative component starts to dominate at low \(Q^2=5 \, \rm GeV^2\).

\begin{figure}[!htb]
	\centering
	\includegraphics[width=\linewidth]{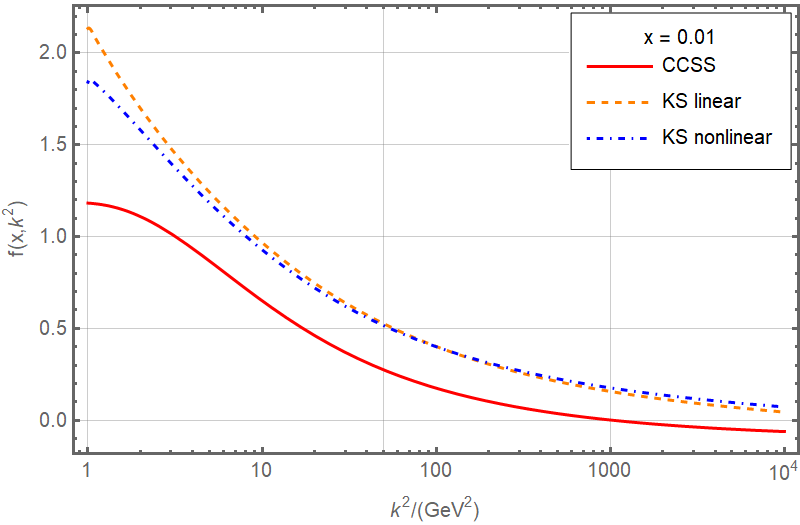}
	\caption{Unintegrated gluon distribution function extracted  from the fit  as a function of the transverse momentum squared  $k^2$ for  fixed \(x = 0.01\). Unintegrated gluon from this work based on CCSS  resummation (red solid) is compared with the other two models, from Ref.~\cite{Kutak:2012rf}, KS linear (dashed orange) and KS non-linear (dashed-dotted blue). }
	\label{fig:gluonplotx0.01}
\end{figure}

\begin{figure}[!htb]
	\centering
	\includegraphics[width=\linewidth]{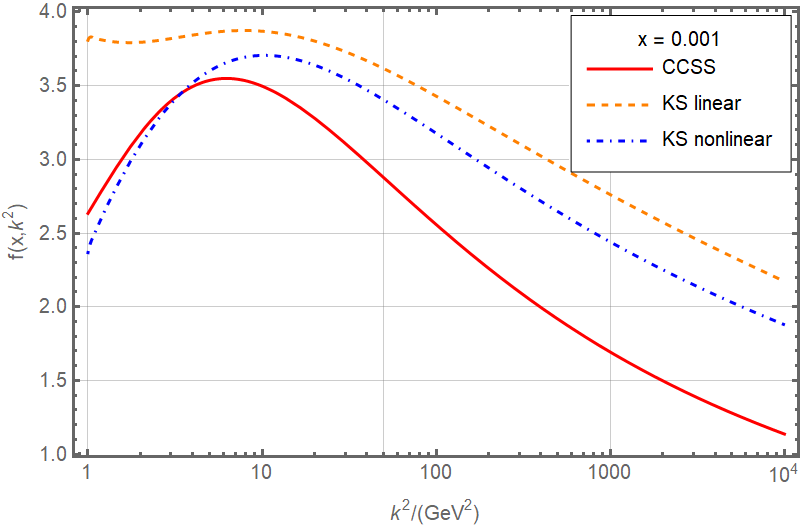}
	\caption{Unintegrated gluon distribution function extracted  from the fit  as a function of the transverse momentum squared  $k^2$ for  fixed \(x = 0.001\). Unintegrated gluon from this work based on CCSS  resummation (red solid) is compared with the other two models, from Ref.~\cite{Kutak:2012rf}, KS linear (dashed orange) and KS non-linear (dashed-dotted blue).}
	\label{fig:gluonplotx0.001}
\end{figure}

In Figs.~\ref{fig:gluonplotx0.01} and \ref{fig:gluonplotx0.001} we show the unintegrated gluon distribution for two fixed values of $x=0.01,0.001$ as a function of  transverse momentum squared $k^2$. The present calculation based on the CCSS resummation is compared with the extracted gluon distribution from \cite{Kutak:2012rf} (called KS for short here) in the linear and non-linear case. The KS calculation is similar in its philosophy to the one presented here and is widely used in the phenomenology, and thus it is interesting to compare it with the current extraction. The KS calculation is based on the LL BFKL with kinematical constraint and includes DGLAP terms, so the main difference is the NLL part of the BFKL and the type of subtractions performed. The overall shape in $k^2$ is  similar for the CCSS and KS calculation. 

\begin{figure}[!htb]
	\centering
	\includegraphics[width=\linewidth]{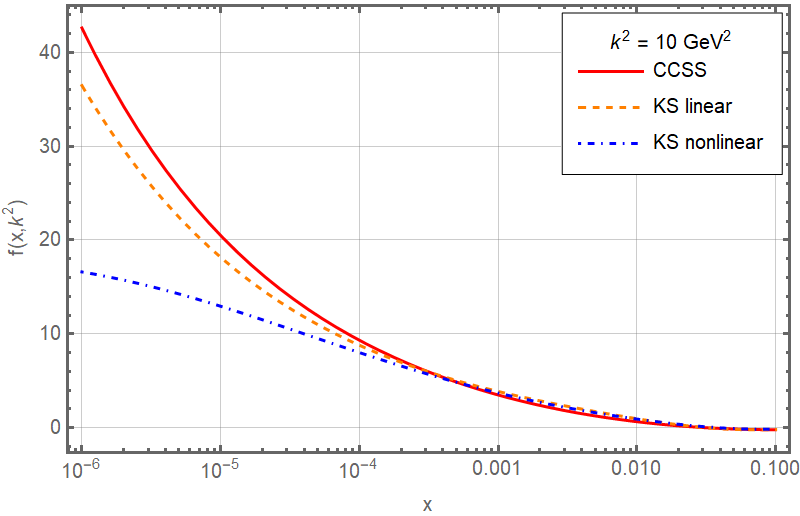}
	\caption{Unintegrated gluon distribution function extracted  from the fit  as a function of $x$ for fixed value of  the transverse momentum squared  \(k^2=10\,\rm GeV^2\). Unintegrated gluon from this work based on CCSS  resummation (red solid) is compared with the other two models, from Ref.~\cite{Kutak:2012rf}, KS linear (dashed orange) and KS non-linear (dashed-dotted blue).}
	\label{fig:gluonplotk210}
\end{figure}

\begin{figure}[!htb]
	\centering
	\includegraphics[width=\linewidth]{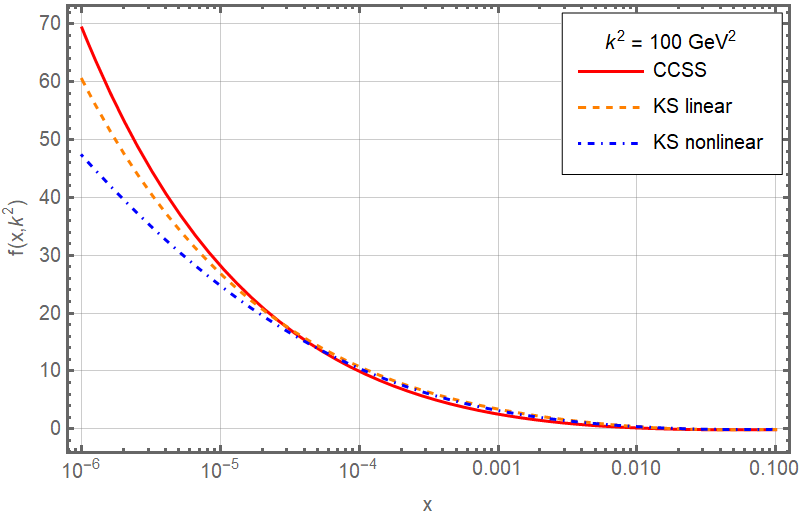}
	\caption{Unintegrated gluon distribution function extracted  from the fit  as a function of $x$ for fixed value of  the transverse momentum squared  \(k^2=100\,\rm GeV^2\). Unintegrated gluon from this work based on CCSS  resummation (red solid) is compared with the other two models, from Ref.~\cite{Kutak:2012rf}, KS linear (dashed orange) and KS non-linear (dashed-dotted blue).}
	\label{fig:gluonplotk2100}
\end{figure}

\begin{figure}[!htb]
	\centering
	\includegraphics[width=\linewidth]{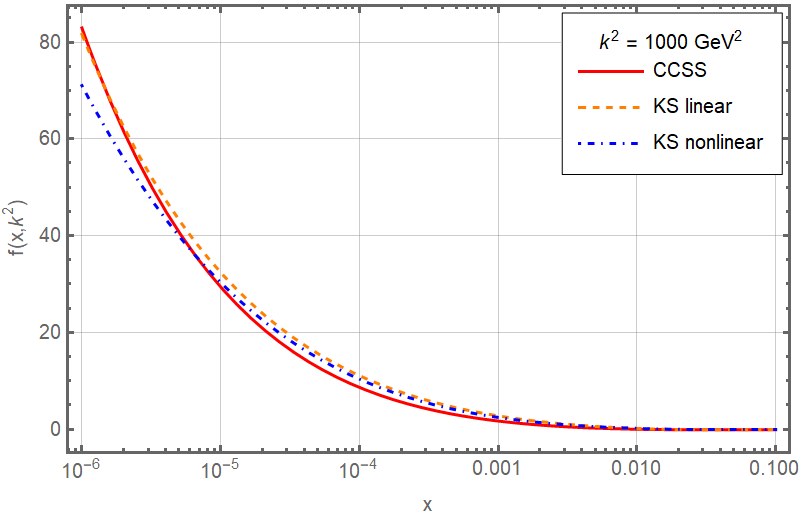}
	\caption{Unintegrated gluon distribution function extracted  from the fit  as a function of $x$ for fixed value of  the transverse momentum squared  \(k^2=1000\,\rm GeV^2\). Unintegrated gluon from this work based on CCSS  resummation (red solid) is compared with the other two models, from Ref.~\cite{Kutak:2012rf}, KS linear (dashed orange) and KS non-linear (dashed-dotted blue).}
	\label{fig:gluonplotk21000}
\end{figure}

In Figs.~\ref{fig:gluonplotk210}-\ref{fig:gluonplotk21000} the unintegrated gluon density is shown as a function of $x$ for fixed values of $k^2=10, 100, 1000 \; \rm GeV^2$, again compared with the KS-linear and KS-nonlinear calculation.  Unsurprisingly, the small $x$ behavior of the CCSS calculation is very close to the KS-linear, whereas both calculations differ substantially from the nonlinear calculation at low $x$ and low $Q^2$ where the saturation corrections are the strongest. There are however some subtle differences between the KS-linear and CCSS calculations which indicate somewhat stronger small $x$ behavior in the CCSS scenario.

\begin{figure}[!htb]
	\centering
	\includegraphics[width=\linewidth]{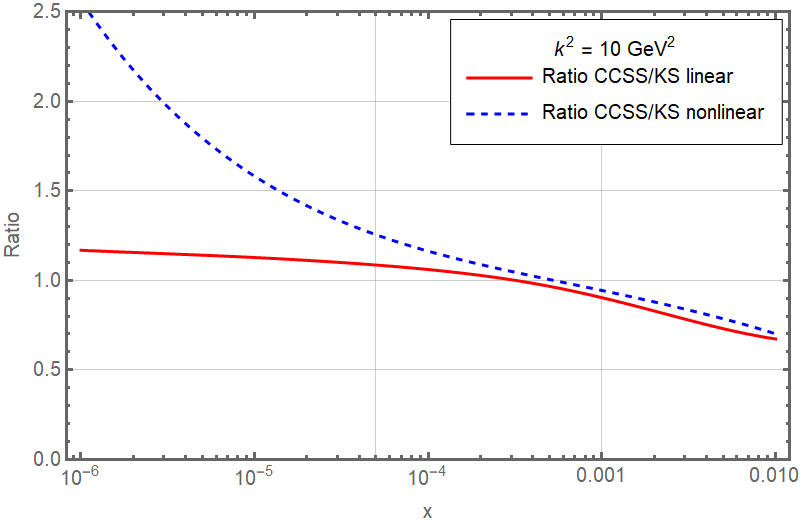}
	\caption{Ratios of unintegrated gluon distributions as a function of $x$ for fixed value of the transverse momentum \(k^2=10 \; \rm GeV^2\). Red solid: ratio of the gluon extracted in this work to the KS linear; blue dashed, ratio of the gluon extracted in this work to the KS non-linear. }
	\label{fig:gluonplotratioQ210}
\end{figure}

\begin{figure}[!htb]
	\centering
	\includegraphics[width=\linewidth]{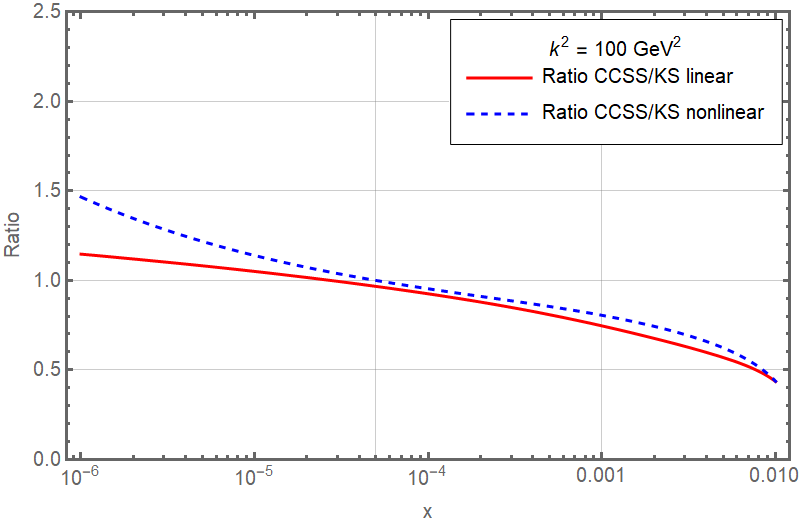}
	\caption{Ratios of unintegrated gluon distributions as a function of $x$ for fixed value of the transverse momentum \(k^2=100 \; \rm GeV^2\). Red solid: ratio of the gluon extracted in this work to the KS linear; blue dashed, ratio of the gluon extracted in this work to the KS non-linear.}
	\label{fig:gluonplotratioQ2100}
\end{figure}

\begin{figure}[!htb]
	\centering
	\includegraphics[width=\linewidth]{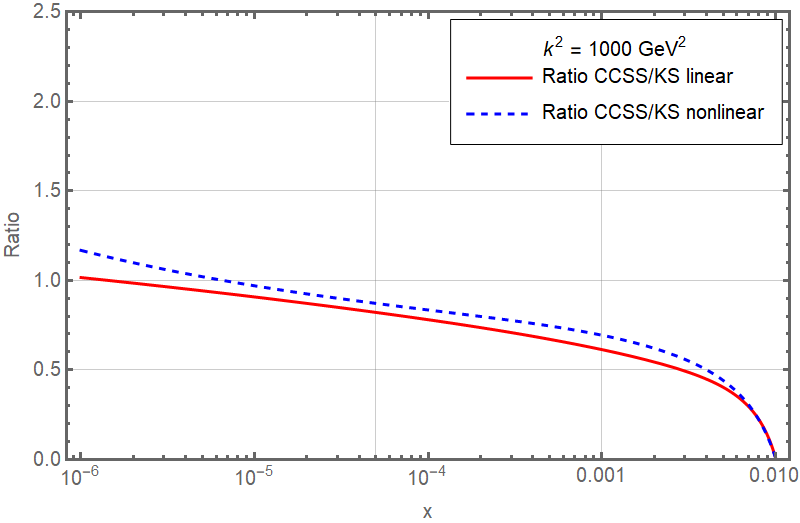}
	\caption{Ratios of unintegrated gluon distributions as a function of $x$ for fixed value of the transverse momentum \(k^2=1000 \; \rm GeV^2\). Red solid: ratio of the gluon extracted in this work to the KS linear; blue dashed, ratio of the gluon extracted in this work to the KS non-linear.}
	\label{fig:gluonplotratioQ21000}
\end{figure}

\begin{figure}[!htb]
	\centering
	\includegraphics[width=\linewidth]{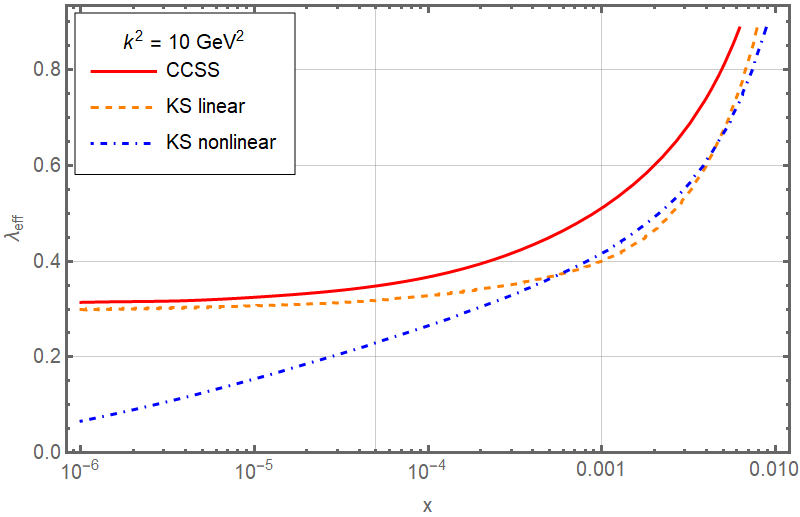}
	\caption{Effective power \(\lambda_{\rm eff}\) from Eq.~\eqref{eq:lambdaeff}  of unintegrated gluon distribution as a function of $x$ for fixed value of the transverse momentum squared \(k^2 = 10 \; \text{GeV}^2\). Solid red, this work; orange dashed, KS linear; blue dashed-dotted KS non-linear.}
	\label{fig:lambda10}
\end{figure}

\begin{figure}[!htb]
	\centering
	\includegraphics[width=\linewidth]{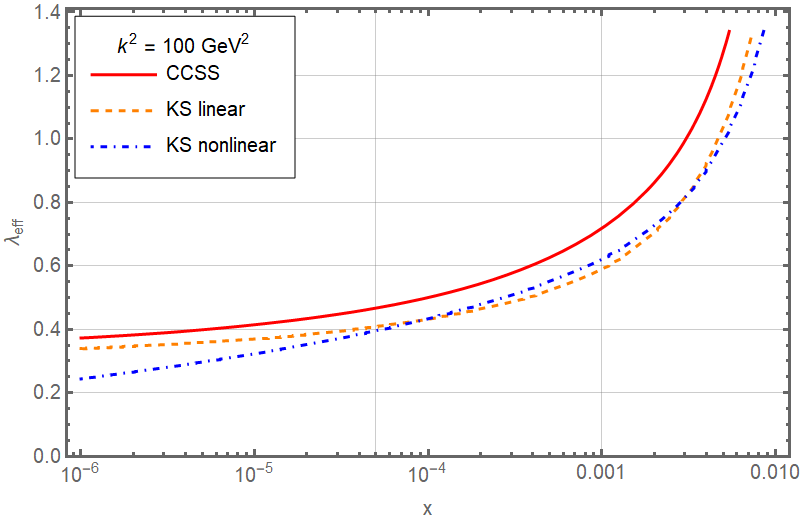}
	\caption{Effective power \(\lambda_{\rm eff}\) from Eq.~\eqref{eq:lambdaeff}  of unintegrated gluon distribution as a function of $x$ for fixed value of the transverse momentum squared \(k^2 = 100 \; \text{GeV}^2\). Solid red, this work; orange dashed, KS linear; blue dashed-dotted KS non-linear.}
	\label{fig:lambda100}
\end{figure}

\begin{figure}[!htb]
	\centering
	\includegraphics[width=\linewidth]{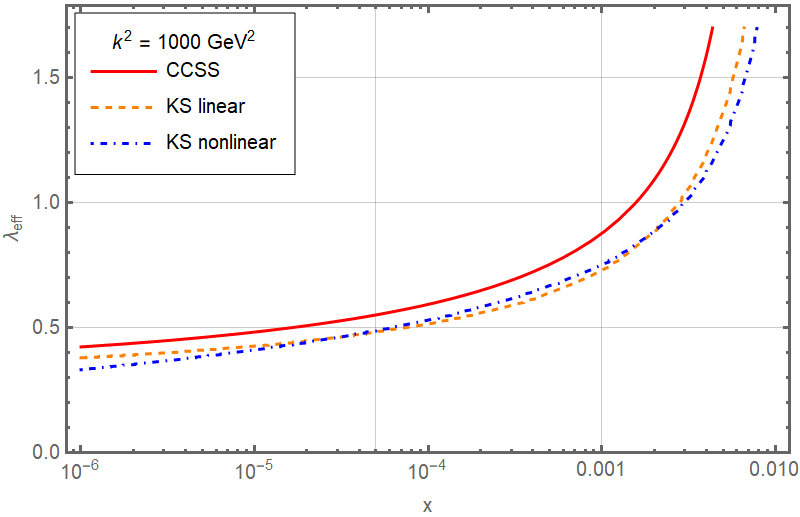}
	\caption{Effective power \(\lambda_{\rm eff}\) from Eq.~\eqref{eq:lambdaeff}  of unintegrated gluon distribution as a function of $x$ for fixed value of the transverse momentum squared \(k^2 = 1000 \; \text{GeV}^2\). Solid red, this work; orange dashed, KS linear; blue dashed-dotted KS non-linear.}
	\label{fig:lambda1000}
\end{figure}

The differences between linear and non-linear scenarios are also visualised by taking the ratios between the calculations shown in Fig.~\ref{fig:gluonplotratioQ210}, \ref{fig:gluonplotratioQ2100}, \ref{fig:gluonplotratioQ21000}.
To be precise the ratios CCSS/KS-linear and CCSS/KS-nonlinear are shown as a function of $x$ for three different values of the momentum squared $k^2=10, 100, 1000 \; \rm GeV^2$. We see that the CCSS/KS-linear ratio is close to unity, for most values of $x$, whereas the ratio to the calculation which includes the nonlinear effects deviates substantially from unity at low $x$ and low $k^2$. Still we see again marked difference, the CCSS calculation tends to have slightly faster rise towards small $x$. The small $x$ behavior is also illustrated in Figs.~\ref{fig:lambda10}, \ref{fig:lambda100} and \ref{fig:lambda1000}, where the effective Pomeron intercept is shown by performing the logarithmic derivative of the unintegrated gluon distribution
\begin{equation}
\lambda_{\rm eff} = \frac{\partial \ln f(x,k^2)}{\partial \ln 1/x} \; .
    \label{eq:lambdaeff}
\end{equation}
It is seen from all plots that the effective power is very close, especially at low $x$ to the linear KS calculation, which is expected, and differs from the non-linear KS, for low $x$ and lowest value of $k^2$. We also see that the power is larger for CCSS calculation than for the linear KS calculation, which is consistent with previous observations. Asymptotically, for low $x$ and low to moderate $k^2$ the value of the power approaches about $0.3$ which is the power observed in HERA data.

\section*{Conclusions and outlook}
\label{sec:conclusions}
In this work we have used  the renormalization group improved small $x$ evolution equation, which resums important terms in the low $x$ region, and used it for the first time to describe the structure function $F_2$ in DIS. Very good description of the experimental data in the low $x<0.01$ region was obtained, with small number of parameters. In addition, the charm structure function was calculated, in two scenarios, where the charm data were used for the fit, and in the second scenario where the structure function $F_2^c$ was not fitted but  obtained instead as  a prediction. In both cases the description of the experimental data was very good and the values of the parameters were very similar. The fit was based on the $k_T$ factorization formula with the  decomposition of the integral in transverse momenta into soft and hard regimes. The unintegrated gluon distribution extracted from this analysis was compared with the previous calculations, based on the similar approach and it was found to be consistent, though with some marked differences. The unintegrated gluon density has somewhat higher power which governs the low $x$ growth as compared to the resummed schemes, which were based on the LL BFKL with kinematical constraint and DGLAP but without the full NLL BFKL terms.  This growth should be  tamed eventually 
by including the non-linear corrections from gluon saturation. 

For the outlook, the current analysis should be potentially expanded to include the DGLAP splitting function up to NNLO, as well as the impact factor up to NLO \cite{Balitsky:2012bs}. Current approach includes LO impact factor, with exact kinematics, which go beyond the lowest order calculation.  In addition, the formulation should include the quarks in the evolution, which could be done by employing the matrix approach to small $x$ evolution as was suggested in \cite{Ciafaloni:2007gf}. We leave these outstanding problems to the future work.

\section*{Acknowledgments}
This work  was  supported in part by the U.S. Department of Energy grant No. DE-SC-0002145 and in part by National Science Centre in Poland, grant 2019/33/B/ST2/02588. We thank Dimitri Colferai and Krzysztof Kutak for discussions and comments.

\bibliographystyle{ieeetr}
\bibliography{mybib}

\end{document}